\documentclass[10pt,a4paper]{article}
\usepackage{geometry}
\usepackage{amsmath,amssymb,slashed,bbm}
\usepackage{setspace}

\renewcommand{\theequation}{\thesection.\arabic{equation}}
\csname@addtoreset\endcsname{equation}{section}

\makeatletter
\pdfoutput=1

\usepackage[T1]{fontenc}
\usepackage{ae}
\usepackage{aecompl}
\usepackage{soul}

\usepackage[sf,bf,small]{titlesec}

\usepackage[small,sf,bf,hang]{caption}

\usepackage[multiple]{footmisc}
\long\def\symbolfootnote[#1]#2{\begingroup%
\def\thefootnote{\fnsymbol{footnote}}\footnote[#1]{#2}\endgroup}

\usepackage{titletoc}
\dottedcontents{section}[6mm]{\smallskip }{6mm}{0pc}
\dottedcontents{subsection}[8mm]{\itshape\small }{6mm}{0pc}
\dottedcontents{subsubsection}[16mm]{\itshape\small }{10mm}{0pc}
\def\tableofcontents{\subsection*{\contentsname}\vspace{-2mm}\@starttoc{toc}}

\usepackage[nosort]{cite}

\usepackage[parsep]{collref}
\usepackage{color}

\usepackage{graphicx}
\usepackage{feynmp}
\usepackage{color}

\DeclareGraphicsExtensions{.pdf}  

\def \bea  {\begin{eqnarray}}
\def \eea  {\end{eqnarray}}

\newcommand{\nn}{\nonumber}

\renewcommand{\u}{\underline}

\begin{document}

\newsavebox{\feynmanrules}
\sbox{\feynmanrules}{
\begin{fmffile}{diagrams} 


\fmfset{thin}{0.6pt}  
\fmfset{dash_len}{4pt}
\fmfset{dot_size}{1thick}
\fmfset{arrow_len}{6pt} 



\begin{fmfgraph*}(80,40)
\fmfkeep{schannel}
\fmfleft{i1,i2}
\fmfright{o1,o2}
\fmf{plain}{i1,v1}
\fmf{plain}{i2,v1}
\fmf{plain,left=0.5,tension=0.4}{v1,v2}
\fmf{plain,right=0.5,tension=0.4}{v1,v2}
\fmf{plain}{v2,o1}
\fmf{plain}{v2,o2}
\fmfdot{v1,v2}
\end{fmfgraph*}

\begin{fmfgraph*}(80,40)
\fmfkeep{tchannel}
\fmfleft{i1,i2}
\fmfright{o1,o2}
\fmf{plain}{i1,v1,o1}
\fmf{plain}{i2,v2,o2}
\fmf{plain,left=0.5,tension=0.4}{v1,v2}
\fmf{plain,right=0.5,tension=0.4}{v1,v2}
\fmfdot{v1,v2}
\end{fmfgraph*}

\begin{fmfgraph*}(80,40)
\fmfkeep{uchannel}
\fmfleft{i1,i2}
\fmfright{o1,o2}
\fmf{plain}{i1,v1}
\fmf{phantom}{v1,o1} 
\fmf{plain}{i2,v2}
\fmf{phantom}{v2,o2} 
\fmf{plain,left=0.5,tension=0.4}{v1,v2}
\fmf{plain,right=0.5,tension=0.4}{v1,v2}
\fmf{plain,tension=0}{v1,o2}
\fmf{plain,tension=0}{v2,o1}
\fmfdot{v1,v2}
\end{fmfgraph*}

\begin{fmfgraph*}(80,40)
\fmfkeep{tadpolesix}
\fmfbottom{i1,o1}
\fmftop{i2,o2}
\fmf{plain}{i1,v1,o1}
\fmf{plain}{i2,v1,o2}
\fmf{plain,right=90,tension=0.8}{v1,v1}
\fmfdot{v1}
\end{fmfgraph*}


\begin{fmfgraph*}(100,36)
\fmfkeep{bubble}
\fmfleft{in}
\fmfright{out}
\fmfdot{v1}
\fmfdot{v2}
\fmf{plain}{in,v1}
\fmf{plain}{v2,out}
\fmf{plain,left,tension=0.6}{v1,v2}
\fmf{plain,right,tension=0.6}{v1,v2}
\end{fmfgraph*}


\begin{fmfgraph*}(100,36)
\fmfkeep{tadpole}
\fmfset{dash_len}{6pt} 
\fmfleft{in,p1}
\fmfright{out,p2}
\fmfdot{c}
\fmf{plain}{in,c}
\fmf{plain}{c,out}
\fmf{plain,right, tension=0.8}{c,c}
\fmf{phantom, tension=0.2}{p1,p2}
\end{fmfgraph*}

\end{fmffile}

}

\makeatletter
\renewcommand{\theequation}{\thesection.\arabic{equation}}
\@addtoreset{equation}{section}
\makeatother

\begin{titlepage}

\vfill
\begin{flushright}
\end{flushright}

\vfill

\begin{center}
   \baselineskip=16pt
   {\Large \bf Worldsheet two and four-point functions at one loop in AdS$_3$ / CFT$_2$}
   \vskip 2cm
    Per Sundin\footnote{E-mail: nidnus.rep@gmail.com}
       \vskip .6cm
             \begin{small}
      		 \textit{
		 Universit\`a di Milano-Bicocca and INFN, Sezione di Milano-Bicocca,  \\
Dipartimento de Fisica, \\
Piazza della Scienza 3, I-20126 Milano, Italy}
             \end{small}\\*[.6cm]
\end{center}

\vfill \begin{center} \textbf{Abstract}\end{center} \begin{quote}
In this note we study worldsheet two and four-point functions at the one-loop level for the type IIA superstring in $AdS_3\times S^3\times M^4$. We first address the regularization ambiguity that appears in the dispersion relation derived from integrability. We demonstrate that only the regulator treating all fields equally respects worldsheet supersymmetry. This is done in an implicit regularization scheme where all divergent terms are collected into master tadpole-type integrals. We then investigate one-loop two-body scattering on the string worldsheet and verify that a recent proposal for the dressing phase reproduces explicit worldsheet computations. All calculations are done in a near-BMN like expansion of the Green-Schwarz superstring equipped with quartic fermions.
\end{quote} \vfill

\end{titlepage}

\tableofcontents


\setcounter{page}{1}
\section{Introduction}
The existence of integrable structures in gauge / string dualities have allowed for a solution of the spectral problem in $\mathcal{N}=4$ SYM. This spectacular achievement is the culmination of a research effort stretching well over the last decade, see \cite{Beisert:2010jr} and references therein.

Intriguingly, not only the maximally supersymmetric $AdS_5 / CFT_4$ is integrable but there seem to exist a plethora of lower dimensional integrable examples \cite{Aharony:2008ug,Babichenko:2009dk,Sorokin:2011rr,Wulff:2014kja}. In this letter we will study certain aspects of one such example, namely the string theory side of $AdS_3 / CFT_2$. The gravitational background is $AdS_3 \times S^3\times M^4$, where $M$ is either $S^3\times S^1$ or $T^4$. The background can furthermore support a combination of non-zero RR and NSNS-fluxes and preserves a total of sixteen supercharges (half that of $AdS_5/CFT_4$) \cite{Cagnazzo:2012se}. The duality was first investigated from the point of view of integrability in \cite{Babichenko:2009dk} where it was found that many of the earlier developed tools could more or less be directly applied. Using well developed methods of the algebraic curve technique and properties of the underlying symmetry, complete sets of Bethe equations for certain of the modes together with proposals for the exact phase have been put forward \cite{Babichenko:2009dk,OhlssonSax:2011ms,Forini:2012bb,Sax:2012jv,Borsato:2012ud,Borsato:2012ss,Borsato:2013qpa,Borsato:2013hoa,Hoare:2013lja,Hoare:2013ida,Hoare:2013pma,Beccaria:2012kb,Lloyd:2013wza,Abbott:2012dd}.

In this letter we want to explore certain quantum properties of the string worldsheet theory, with zero NSNS-flux, building on earlier results of \cite{Sundin:2013ypa,Sundin:2012gc,Abbott:2011xp}. The underlying symmetry algebra constrain the dispersion relation of single excitation magnons up to an overall function $h(\lambda)$ which enters as the coupling constant in the integrable picture. Similarly to the type IIA string in $AdS_4\times\mathbbm{CP}_3$ a regularization ambiguity arises when one tries to write it in terms of the string coupling $g=\sqrt\lambda / 4\pi$. The ambiguity is traced back to the heavy worldsheet modes of the theory which in the exact solution is a composite state of two lighter ones. This leads to two natural regularization schemes which gives different predictions for $h(\lambda)$. By looking at two-point functions built out of the light modes we show that only one regulator is consistent with the symmetries of the problem resulting in a non-zero one-loop correction. This has earlier been argued for in $AdS_4/CFT_3$ by looking at methods developed for the supersymmetric two dimensional kink \cite{LopezArcos:2012gb}.

We then turn to investigate a recent proposal for the one-loop phase factor in the strict $AdS_3\times S^3\times T^4$ case. The one-loop phase was first proposed in \cite{Beccaria:2012kb} but later augmented with additional Kronecker-delta terms \cite{Borsato:2013hoa, Abbott:2013ixa}. In \cite{Sundin:2013ypa} the one-loop phase was probed using the Near Flat Space (NFS) string, a limit of the string theory where the right moving sector of the theory is boosted \cite{Maldacena:2006rv}. The resulting theory is vastly simplified compared to the full BMN string but nevertheless maintain some non-trivial features. However, for the NFS the additional terms in the modified phase appears at subleading order in perturbation theory and could not be verified in the setting of \cite{Sundin:2013ypa}. Here we redo the computation using the full (near)-BMN string. By computing four-point functions on the worldsheet ($2\rightarrow 2 $ scattering) we verify that the modified phase indeed agrees with explicit string theory calculations.

The outline of the paper is as follows: We begin in section \ref{sec:setup} with a quick review of the GS-string expanded to quartic order in fermions. We also discuss light-cone and kappa-gauge fixing together with the BMN spectrum and its corresponding expansion in large $g$. We then turn to the investigation of loop-corrected two-point functions in section \ref{sec:twopoint}. In section \ref{sec:fourpoint} we verify that the phase of \cite{Borsato:2013hoa, Abbott:2013ixa} indeed reproduces explicit string data. We end the paper with a short summary and discussion.

Throughout the paper we will be very brief in technical detail. For the interested reader the quartic GS string is derived in \cite{Wulff:2013kga} by Wulff (see also \cite{Wulff:2014kja}), lengthy discussions about loop corrected two-point functions can be found in \cite{Sundin:2012gc,Abbott:2011xp} and similar discussions for four-point functions are found in \cite{Sundin:2013ypa}.
\section{Setup}
\label{sec:setup}
For a general type IIA supergravity background, the Green-Schwarz string takes the form
\bea
\label{eq:action}
S=-g\int_\Sigma \big(\frac{1}{2}*\, E^a\, E^b \eta_{ab}-B\big)
\eea
where $g$ is proportional to the string tension, $E^a=E^a(X,\Theta)$ is the ten dimensional supervielbein and $B$ is the NSNS two-form potential projected upon the string worldsheet $\Sigma$. Wedge products are left implicit and the * denotes Hodge-dual defined with respect to a two-dimensional worldsheet metric $\gamma^{ij}=\sqrt{-h}h^{ij}$. Expanding the action in fermions as $\mathcal{L}=\mathcal{L}^{(0)}+\mathcal{L}^{(2)}+\dots$, we have to quartic order with constant Dilaton and zero NSNS-flux \cite{Cvetic:1999zs, Wulff:2013kga},
\bea
&& \mathcal{L}^{(0)}=\frac{1}{2}* e^a e^b \eta_{ab}, \\ \nn
&& \mathcal{L}^{(2)}=\frac{i}{2}* e^a \Theta \Gamma_a \mathcal{D} \Theta-\frac{i}{2}\Theta \slashed e \Gamma_{11}\mathcal D\Theta,
\eea
\begin{eqnarray} \nn
\lefteqn{\mathcal L^{(4)}=}
\nonumber\\
&&{}
-\frac{1}{8}\Theta\Gamma^a*\mathcal D\Theta\,\Theta\Gamma_a\mathcal D\Theta
+\frac{1}{8}\Theta\Gamma^a\mathcal D\Theta\,\Theta\Gamma_a\Gamma_{11}\mathcal D\Theta
+\frac{i}{24}*e^a\,\Theta\Gamma_a\mathcal M\mathcal D\Theta
-\frac{i}{24}e^a\,\Theta\Gamma_a\Gamma_{11}\mathcal M\mathcal D\Theta
\nonumber\\
&&{}
+\frac{i}{3\cdot64}*e^ae^b\,\Theta\Gamma_a(M+\tilde M)S\Gamma_b\Theta
-\frac{i}{3\cdot64}e^ae^b\,\Theta\Gamma_a\Gamma_{11}(M+\tilde M)S\Gamma_b\Theta
\nonumber\\
&&{}
+\frac{1}{3\cdot64}[*e^ce^d\,\Theta\Gamma_c{}^{ab}\Theta-e^ce^d\,\Theta\Gamma_c{}^{ab}\Gamma_{11}\Theta]\,(3\Theta\Gamma_dU_{ab}\Theta
-2\Theta\Gamma_aU_{bd}\Theta)
\nonumber\\
&&{}
-\frac{1}{3\cdot64}[*e^ce^d\,\Theta\Gamma_c{}^{ab}\Gamma_{11}\Theta-e^ce^d\,\Theta\Gamma_c{}^{ab}\Theta]\,(3\Theta\Gamma_d\Gamma_{11}U_{ab}\Theta+2\Theta\Gamma_a\Gamma_{11}U_{bd}\Theta)\,.
\nn
\end{eqnarray}
where the long derivative is the Killing spinor operator,
\bea
\nn
\mathcal{D}=\big(d - \frac{1}{4}\slashed \Omega+\frac{1}{8} S \slashed e \big)\Theta\quad \text{with}\quad  S=\frac{1}{2}F^{(2)}_{AB}\Gamma^{AB}\Gamma_{11}+\frac{1}{4!}F^{(4)}_{ABCD} \Gamma^{ABCD}.
\eea
Following the notation of \cite{Wulff:2013kga} we use
\begin{eqnarray}
\mathcal M^{\u\alpha}{}_{\u\beta}&=&
M^{\u\alpha}{}_{\u\beta}
+\tilde M^{\u\alpha}{}_{\u\beta}
+\frac{i}{8}(S\Gamma^a\Theta)^{\u\alpha}\,(\Theta\Gamma_a)_{\u\beta}
-\frac{i}{16}(\Gamma^{ab}\Theta)^{\u\alpha}\,(\Theta\Gamma_aS\Gamma_b)_{\u\beta}
\nonumber\\
M^{\u\alpha}{}_{\u\beta}&=&
\frac12\Theta T\Theta\,\delta^{\u\alpha}_{\u\beta}
-\frac12\Theta\Gamma_{11}T\Theta\,(\Gamma_{11})^{\u\alpha}{}_{\u\beta}
+\Theta^{\u\alpha}\, (T\Theta)_{\u\beta}
+(\Gamma^aT\Theta)^{\u\alpha}\,(\Theta\Gamma_a)_{\u\beta}
\label{eq:M}
\end{eqnarray}
with $\tilde M=\Gamma_{11}M\Gamma_{11}$. The two matrices $T$ and $U$ appear in conditions for supersymmetry preserved by the background and are explicitly given by
\begin{eqnarray}
T=\frac{i}{16}\Gamma_aS\Gamma^a,\qquad
U_{ab}=
\frac{1}{4}\nabla_{[a}S\Gamma_{b]}
-\frac{1}{4}R_{abcd}\,\Gamma^{cd}
+\frac{1}{32}S\Gamma_{[a}S\Gamma_{b]}
\end{eqnarray}
The $AdS_3\times S^3\times M^4$  backgrounds we will study in this letter can be unified using a smooth parameter $\alpha \in [0,1]$ which parameterize the relative radii between the two $S^3$-spheres. At the boundary points, $\alpha=0$ and $\alpha=1$, the background geometry decompactify to $AdS_3\times S^3\times T^4$. This implies that the string Lagrangian is in fact parameterized by two free parameters; the string coupling $g$ (which involves the $AdS$-curvature) and the parameter $\alpha$. Furthermore, the 4-form RR-flux is written in terms of this parameter via \cite{Babichenko:2009dk},
\bea
S=-4\Gamma^{0129}\big(1-\mathcal{P}\big),\qquad \mathcal{P}=\frac{1}{2}\big(1+\sqrt\alpha\,\Gamma^{012345}+\sqrt{1-\alpha}\,\Gamma^{012678}\big)
\eea
where $\mathcal{P}$ is a projector that singles out the 16 supersymmetries preserved by the background. For the metric we will employ the following coordinates for the line segment,
\bea
\nn
&& ds^2=
-\left(\frac{1+\frac12 |y_1|^2}{1-\frac12 |y_1|^2}\right)^2dt^2+\frac{2}{\big(1-\frac12 |y_1|^2\big)^2} |dy_1|^2
+\left(\frac{1-\frac{\alpha}{2} |y_2|^2}{1+\frac{\alpha}{2}|y_2|^2}\right)^2d\varphi_1^2+\frac{2}{\big(1+\frac{\alpha}{2} |y_2|^2\big)^2} |dy_2|^2 \\ \nn
&&
+\left(\frac{1-\frac{1-\alpha}{2} |y_3|^2}{1+\frac{1-\alpha}{2}|y_3|^2}\right)^2d\varphi_2^2
 +\frac{2}{\big(1+\frac{1-\alpha}{2} |y_3|^2\big)^2} |dy_3|^2+dx_9^2
\eea
Equipped with the string Lagrangian and the parameterization of background fluxes and metric we are in position to expand the action in a strong coupling, or BMN like, expansion.
\subsection*{BMN expansion}
We will utilize light-cone pairs that combine the $AdS$ time coordinate with two angles from $S^3\times S^3$,
\bea x^\pm=\frac{1}{2}\left(t+\sqrt\alpha \,d\varphi_1+\sqrt{1-\alpha}\,d\varphi_2\right),\qquad
\Gamma^\pm=\Gamma^0\pm \sqrt\alpha \,\Gamma^5+\sqrt{1-\alpha}\,\Gamma^8,
\eea
and then use the worldsheet 2D Weyl and diffeomorphism invariance together with local $\kappa$-symmetry to fix bosonic and fermionic light-cone gauges as\footnote{ For explicit representations of $\Gamma$-matrices and such we point to \cite{Rughoonauth:2012qd,Sundin:2010aea}. }
\bea
\label{eq:lc-gauge}
x^+=\tau,\qquad p_-=\text{constant},\qquad \gamma_{ij}=\eta_{ij}+\mathcal{O}(g^{-1/2}),\qquad \Gamma^+\Theta=0
\eea
where the higher-order terms for the metric are fixed via the equations of motion for $x^-$. The light-cone pair parameterize a null geodesic which is the solution we will perform a semi-classical expansion around. Scaling the transverse bosonic and fermionic string coordinates with $g^{-1/2}$ we can expand the action (\ref{eq:action}) as \cite{Berenstein:2002jq,Callan:2004uv}
\bea
\label{eq:bmn-string}
\mathcal{L}=\mathcal{L}_2+g^{-1/2}\mathcal{L}_3+g^{-1}\mathcal{L}_4+\dots
\eea
where the subscript denotes number of transverse coordinates and we dropped the fermionic superscript. Effectively this results in a 2D field theory with a free, quadratic, Lagrangian given by $4_B+4_F$ complex fields
\bea
\mathcal{L}_2=i \bar \chi_+^a  \partial_- \chi_+^a+i\bar \chi_-^a \partial_+ \chi_-^a+|\partial y_a|^2-m_a \big(\bar\chi^a_+ \chi^a_- +\bar\chi^a_- \chi^a_+\big)-m_a^2 |y_a|^2
\eea
where $m_a=\big(1,\alpha,1-\alpha,0\big),\,\partial_\pm=\partial_0\pm \partial_1$ and the subleading terms should be thought of as interaction terms giving rise to non-trivial Feynman topologies on the string worldsheet. In the remainder of this paper we will work with two specific values for $\alpha$. In the analysis of two-point functions we will work with $\alpha=\frac{1}{2}$ where the "light" coordinates $(y_2,\chi^2)$ and $(y_3,\chi^3)$ have the same mass. In the subsequent section of the paper, where we consider $2\rightarrow 2$ scatterings on the string worldsheet, we will work in the decompactifying case $\alpha=1$. In this limit 1 and 2 coordinates have unit mass while the remaining fields are massless.

After the light-cone gauge fixing we have three conserved U(1) charges,
\[ \label{tab:charges}
\begin{tabular}{c|cccc|cccc}
 & $y_1$& $y_2$ & $y_3$ & $y_4$ & $\chi^1_\pm$ & $\chi^2_\pm$ & $\chi^3_\pm$ & $\chi^4_\pm$ \tabularnewline \hline
mass & $1$ & $\alpha$ & $1-\alpha$ & $0$ & $1$ & $\alpha$ & $1-\alpha$ &$0$ \tabularnewline \hline
$U(1)_{AdS}$ & -1 & 0 & 0 & 0& $-1/2$ & $1/2$ & $1/2$ & $1/2$ \tabularnewline
$U(1)_{S^3_+}$ & 0 & -1 & 0 &0&  $1/2$ & $-1/2$ & $1/2$ & $1/2$  \tabularnewline
$U(1)_{S^3_-}$ & 0 & 0 & -1 & 0& $1/2$ & $1/2$ & $-1/2$ & $1/2$ \tabularnewline \hline
\end{tabular}
\]
Furthermore, the first non-trivial piece of the Lagrangian start at cubic order in fields and is explicitly given by
\bea
&& \mathcal{L}_3=\frac{1}{\sqrt{2}}\sqrt{(1-\alpha)\alpha}\Big[ \alpha\big(\chi^1_- \chi^3_-+\chi^4_- \bar\chi^2_--\chi_+^1\chi^3_++\chi_+^4 \bar\chi_+^2\big)y_2
\\ \nn
&& +(1-\alpha)\big(-i \chi^1_- \chi^2_- +i \chi^4_-\bar\chi^3_-+i\chi^1_+ \chi^2_++i\chi^4_+ \bar\chi^3_+\big)y_3 \\ \nn
&&+\big(-i\chi^3_-\chi^1_+-i\chi^4_+\bar\chi^2_-\big)\partial_+ y_2+\big(-i\chi^1_-\chi^3_+-i\chi^4_-\bar\chi^2_+\big)\partial_- y_2
\\ \nn
&&+\big(-\chi^2_-\chi^1_++\chi^4_+\bar\chi^3_-\big)\partial_+ y_3+\big(-\chi^1_-\chi^2_++\chi^4_-\bar\chi^3_+\big)\partial_- y_3 \\ \nn
&& +2\big(-\chi^2_- \chi^3_+-\chi^3_-\chi^2_+\big)y'_1+2\big(i\chi^2_+\bar\chi^2_--i\chi^3_+\bar\chi^3_-\big)\dot y_4
-2i\big(\alpha |y_2|^2-(1-\alpha)|y_3|^2\big) \dot y_4+h.c\Big]
\eea
Here we note something interesting, in the strict $\alpha=0$ or $\alpha=1$ limit, where the geometry is $AdS_3\times S^3\times T^4$, we see that the cubic Lagrangian vanishes. In fact, this is a general feature and at higher orders in perturbation theory all odd-numbered vertexes vanishes. We will not present any of the higher order terms of the Lagrangian, but the interested reader can find them in \cite{Rughoonauth:2012qd}.

As a concluding remark for this section we note that the theory is symmetric under a permutation of the two $S^3$. At the level of the string worldsheet, this permutation is realized via the following $\mathbbm{Z}_2$ worldsheet transformations,
\bea \nn
\alpha\rightarrow 1-\alpha,\qquad \chi^2_\pm \rightarrow i \chi^3_\pm,\qquad \chi^3_\pm \rightarrow -i \chi^2_\pm,\qquad \chi^4_\pm\rightarrow -\chi^4_\pm,\qquad y_2\leftrightarrow y_3,\qquad y_4\rightarrow -y_4
\eea
\section{Two-point functions at one loop}
\label{sec:twopoint}
As we discussed in the introduction, strings in both $AdS_4\times \mathbbm{CP}_3$ and $AdS_3\times S^3\times S^3\times S^1$ exhibit regularization ambiguities for loop-corrected propagators. This manifests itself in ambiguous finite terms which results in a regularization dependent dispersion relation (or, equivalently a one-loop normalization of the coupling $g$). This is naturally a somewhat unsatisfactory situation since the dispersion relation is a physical quantity. On the other hand, the occurrence of finite ambiguities in loop calculations is definitely not something new and it is known that the difference between divergent loop integrals of the same superficial degree of divergence often result in finite, regularization dependent, terms. Thus, some care is needed when computing the divergent integrals.

The strategy we will use here is known as implicit regularization where the key idea is to separate UV divergent integrals into a divergent master-class integral plus finite integrals of different types. Once this is done a regulator is picked which respects the underlying symmetries of the problem \cite{Hiller:2005uw,Jackiw:1999qq,BaetaScarpelli:2000zs,Sampaio:2002ii,Carneiro:2003id,Souza:2005vf}. For the case at hand this means the regulator should leave the two dimensional worldsheet QFT supersymmetric and respect the underlying residual global symmetry (left after fixing the light-cone gauge and expanding around the BMN vacuum).\footnote{The worldsheet supersymmetry originates from the $\kappa$-gauge fixing and has no connection with the worldsheet supersymmetry in the RNS formulation of the superstring.} In other words,
\begin{itemize}
\item  worldsheet susy demands\footnote{Since $y_i$ and $\chi^i$ should be superpartners on the worldsheet each pair need to have the same mass renormalization.} $\langle \bar y_i\, y_i \rangle \vert_{p_1\rightarrow 0}= \langle \bar \chi^i\, \chi^i \rangle\vert_{p_1\rightarrow 0}\qquad \qquad(i=2,3)$
\item  while underlying global symmetry implies the same regulator for 2 and 3 particles.
\end{itemize}
For the actual calculation we encounter two types of diagrams\footnote{Proper tadpoles built out of three-vertexes are trivially zero.}
\begin{align}
\label{bubble}
\parbox[top][0.8in][c]{1.5in}{\fmfreuse{bubble}}
\quad + \quad
\parbox[top][0.8in][c]{1.5in}{\fmfreuse{tadpole}}
\end{align}
giving rise to standard integrals of the form
\bea \nn
B^{rs}(a,b)=\frac{1}{(2\pi)^2}\int d^2k \frac{k_+^r k_-^s}{\big(k^2-m_a^2\big)\big((k-p)^2-m_b^2\big)},\qquad T^{rs}(a)=\frac{1}{(2\pi)^2}\int d^2k \frac{k_+^r k_-^s}{k^2-m_a^2}
\eea
The naive way to evaluate the one-loop correction is to simply compute the combinatoric pre-factors for each diagram and then evaluate the actual integrals. The final result is then a sum of bubble and tadpole contributions which gives a value for the one-loop corrected propagator. However, as advocated, this is not the correct way as we will demonstrate below.

Throughout this section we will work with two distinct regularization schemes. The first scheme we will denote worldsheet-regulator (WS) where all fields are treated equally and cutoff a the same momentum (equivalent to standard dimensional regularization). This scheme is natural from the worldsheet point of view where each field is simply one of the eight transverse degrees of freedom left after the light-cone gauge fixing. The second cutoff is denoted algebraic curve (AC) and is natural in the exact solution where the heavy mode is a composite state built out of two light ones \cite{Gromov:2008bz,Gromov:2008fy}. Here twice the momentum cutoff is used for the heavy mode as compared to the two lighter ones.\footnote{ When $\alpha=\frac{1}{2}$ this follows from $\omega_1(2p)=2\omega_{2,3}(p)$.}

We will evaluate the two-point functions both in the standard way, where we sum divergent tad and bubble-integrals, and in the implicit regularization-scheme. Using textbook QFT methods we find using WS-regulator (for simplicity we put $\alpha=\frac{1}{2}$ in this section),
\bea
\label{eq:naive-WS-2point}
\langle \bar y_i\, y_i \rangle^{(1)}=-\frac{i}{\pi\,g}\log 2\,p_1^2,\qquad \langle \bar \chi_\pm^i\, \chi_\pm^i \rangle^{(1)}=-\frac{i}{16\pi\,g}-\frac{i}{\pi \,g}\big(\frac{1}{8}+\log 2\big)p_1^2
\eea
where $i=2,3$. The bosonic propagator has earlier been computed in \cite{Sundin:2012gc} and the superscript on the propagator denotes one-loop piece. Evidentally worldsheet supersymmetry is broken since the mass renormalization for the superpartners is different. Furthermore, since the bubble integrals generically have particles of different masses in the loop its not clear how to implement the AC-regulator for divergent diagrams.\footnote{ However, one could try to implement it by regulating the bubbles as light-modes and the tadpoles according to scheme. This was for example done in \cite{Sundin:2012gc,Abbott:2011xp}. However, there is no physical motivation for this scheme (and one can verify that it also breaks worldsheet supersymmetry) so we will only present results for the AC-regulator when it can be properly implemented.} For this reason we will only employ the AC-regulator in implicit regularization (where the regulator can be unambiguously implemented).

We will now redo the above computation using implicit regularization. The first step will be to isolate the divergences from bubbles in terms of tadpole integrals, written without any dependence on the external momentum. The divergent bubble integrals have $(r,s)=(1,1),(1,2),(2,1)$ power of $k_\pm$ in the numerator, and using
\bea
B^{rs}(a,b)=\frac{1}{(2\pi)^2}\int d^2 k \frac{k_+^{r-1}k_-^{s-1}}{(k-p)^2-m_b^2}+m_a^2 B^{r-1\,s-1}(a,b)
\eea
we see that we can isolate the divergency in the first term. For this term it is tempting to shift the loop integration variable as $k\rightarrow k+p$ and perform the integral. However, this still implies that we sum different types of integrals, which is what we are trying to avoid. If we instead use a second algebraic identity
\bea
\frac{1}{(k-p)^2-m^2}=\frac{1}{k^2-m^2}-\frac{p^2-2 p\cdot k}{\big(k^2-m^2\big)\big((p-k)^2-m^2\big)}
\eea
the UV divergency is isolated in a proper tadpole-type integral. This procedure can be repeated until all bubble integrals are manifestly finite. Once this is done we are in position to evaluate the two-point functions. Furthermore, all divergent terms are isolated in proper tadpole-type integrals so its now clear how to implement the AC regulator. Using both schemes we find
\bea
&& \langle \bar y_i\, y_i \rangle_{WS}^{(1)}=-\frac{i}{\pi\,g}\log 2\,p_1^2,\qquad \langle \bar \chi_\pm^i\, \chi_\pm^i \rangle_{WS}^{(1)}=-\frac{i}{\pi\,g}\log 2\,p_1^2, \\ \nn
&& \langle \bar y_i\, y_i \rangle_{AC}^{(1)}=\frac{i}{8\pi\,g}\big(\frac{3}{8}\log 2-\log 2\,p_1^2\big)-\frac{i}{16\pi\,g}\log 2\,p_1 \sqrt{1+4p_1^2},\qquad \langle \bar \chi_\pm^i\, \chi_\pm^i \rangle_{AC}^{(1)}=\frac{i}{16\pi\,g}\log 2
\eea
and thus only the WS-regulator, where we treat all fields equally, leaves worldsheet supersymmetry intact.\footnote{ We have ignored constant $\Lambda^2$ terms from tadpole integrals. These are a defect of the hard cutoff method and are not present in, say, dimensional regularization (they appear also in, for example, $AdS_5 \times S^5$). If the reader feels awkward about throwing away the $\Lambda^2$ terms by hand, one can modify the dimensional regularization scheme to incorporate a composite heavy mode, see \cite{Sundin:2012gc} for details. Using this everything is manifestly finite and we reproduce the findings in the main text.} It is interesting to note that when we implement the AC regulator properly, using implicit regularization, we are unable to remove the $-\log 2\, p_1^2$ term in the bosonic propagator. What is worse, for this regulator we get an explicit contradiction with integrability since the dispersion relation is not of a non-relativistic sine-square root type.

While we have not checked this here, it is plausible that we would find the same result for $AdS_4\times \mathbbm{CP}_3$. In fact, this is already hinted at from the analysis in \cite{LopezArcos:2012gb} where methods from the supersymmetric 2D kink was used to argue for a non-zero one-loop correction.
\section{Four-point functions at one loop}
\label{sec:fourpoint}
We now turn to investigate four point functions at the one loop level. As for earlier incarnations of $AdS / CFT$, the underlying integrable structures can be used to determine the dispersion relation and 2-body S-matrix (modulo an overall phase factor). Perhaps somewhat surprisingly the phase turned out to be the same in both $AdS_5 / CFT_4$ and $AdS_4 /CFT_4$ and is usually denoted $\sigma_{BES}$ \cite{Beisert:2006ez,Gromov:2008qe,Gromov:2008bz}. For $AdS_3/CFT_2$ the phase is the BOSST phase which is different (but related) to the standard BES-phase. Furthermore, for the less studied $AdS_2 / CFT_1$ the BES-phase again makes an entrance, albeit in different overall power than earlier examples \cite{Abbott:2013kka}.\footnote{ There is, as of yet, no information beyond the one loop level.}

The tree and one-loop components of BES in strong coupling expansion are the well known AFS and HL-phases \cite{Arutyunov:2004vx, Hernandez:2006tk}. Both BES and BOSST share the AFS phase but they differ at the one-loop level. Writing $\sigma=e^{i \theta}$, the phase structure for known (and integrable) $AdS / CFT$ summarizes to one-loop (at strong coupling) as:
\begin{equation}\label{eq:list-of-Theta}\begin{aligned}
\qquad\qquad AdS_{5}\times S^{5}: & \vphantom{\frac{1}{1_{1}^{1}}} & \theta=\;
& 2 h \theta_{\text{AFS}}+2\theta_{\text{HL}}\negthickspace\negthickspace & \negthickspace\negthickspace & =2\theta_{\text{BES}}(h)\qquad\qquad\\
AdS_{4}\times CP^{3}: &  &
& h\theta_{\text{AFS}}+\theta_{\text{HL}}+\ldots & \negthickspace\negthickspace & =\theta_{\text{BES}}(h)\\
AdS_{3}\times S^{3}\times T^{4}: &  &
& 2h\theta_{\text{AFS}}+\begin{cases}
2\theta_{LL}+\ldots\\
2\tilde{\theta}_{LR}+\ldots
\end{cases} & \negthickspace\negthickspace & =\begin{cases}
2\theta_{\text{BOSST}}\\
2\tilde{\theta}_{\text{BOSST}}
\end{cases}\\
AdS_{2}\times S^{2}\times T^{6}: &  &
& 4 h \theta_{\text{AFS}}+2\theta_{\text{HL}}+\ldots & \negthickspace\negthickspace & =2\theta_{\text{BES}}(2h).
\end{aligned}\end{equation}
In \cite{Sundin:2013ypa} both LL and LR pieces of BOSST were explicitly verified in the so called Near-Flat Space (NFS) limit where the worldsheet right moving sector is boosted as $\partial_- \sim g^{1/2}$. This limit is nice since being a truncation of the full BMN string, it nevertheless maintains non-trivial interactions. However, being a truncation, not all terms in LL and LR could be verified. In particular, the first proposals for $LL$ and $LR$ came in \cite{Beccaria:2012kb} but these were later modified in \cite{Borsato:2013hoa,Abbott:2013ixa} with additional Kronecker-delta terms (coming from boundary terms and symmetrization arguments). These additional terms are invisible in the strict NFS and it is necessary to consider the full BMN-string to explicitly see them.

In this section we will explicitly check these one-loop terms. However, we have two problems. First, the general structure of the full BMN-string is fairly intricate and involves vertexes up to sixth order in fields. For this reason we will only study the strict $\alpha=1$ case where a $S^3\times S^1$-factor decompactifies to $T^4$. This has the upshot that we only have to consider four and six vertexes in our diagrammatic expansion (which still is fairly involved). Second, and more important, the light-cone gauge fixed BMN string fails to be manifestly UV finite on the string worldsheet \cite{Engelund:2013fja, Abbott:2013kka}. Luckily however, the UV divergencies only appear in "diagonal" processes $y_i\, y_i \rightarrow y_i\, y_i$ and we can still explicitly verify the complete LL and LR factors by computing\footnote{The fact that the inconsistencies with the light-cone gauge only affects diagonal processes might hint that the solution has to do with the Virasoro-constraint, which may need corrections at the one-loop level. Another possible explanation could be that the S-matrix is, in fact, gauge dependent. It would be interesting to compute the two-loop corrected dispersion relation utilizing the full BMN-string. However, due to the complexity of the computation this has not even been performed for the more symmetric $AdS_5\times S^5$-string.}
\bea \label{eq:22scatterings}
&& i \, \mathcal{A}^{(1)}\Big[y_1(p)\, y_2(q)\rightarrow  y_1(p)\,y_2(q)\Big] = \tilde\theta_{LR}(p,q) +\text{real terms}\\ \nn
&& i \, \mathcal{A}^{(1)}\Big[y_1(p)\, \bar y_2(q)\rightarrow  y_1(p)\,\bar y_2(q)\Big] = \theta_{LL}(p,q)+\text{real terms}
\eea
where the subscript on $\mathcal{A}$ denotes one-loop component of the amplitude. The real terms are not very interesting since they are completely determined by tree-level amplitudes via the optical theorem. They are furthermore also completely constrained by the symmetries of the problem.

The two amplitudes (\ref{eq:22scatterings}) boils down to the summation of four distinct diagrams: Three four-vertex topologies,
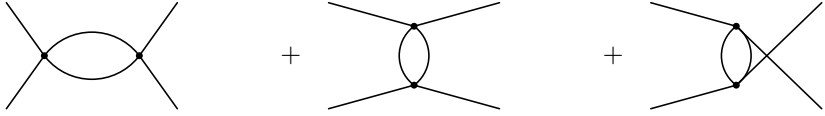
\begin{align} \nn
\parbox[top][0.8in][c]{1.5in}{\fmfreuse{schannel}}+\parbox[top][0.8in][c]{1.5in}{\fmfreuse{tchannel}}+\parbox[top][0.8in][c]{1.5in}{\fmfreuse{uchannel}}
\end{align}
and one six-vertex tadpole topology
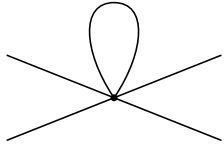
\begin{align} \nn
\parbox[top][0.8in][c]{1.5in}{\fmfreuse{tadpolesix}}
\end{align}
Each class of diagrams is divergent but summed together the resulting amplitude is manifestly IR and UV finite.\footnote{ We have evaluated the diagrams using standard methods and have checked that implicit regularization gives the same answer.} After a fairly involved computation we find for the first amplitude in (\ref{eq:22scatterings})
\bea
&&  i \, \text{Im }\mathcal{A}^{(1)}\Big[y_1(p)\, y_2(q)\rightarrow  y_1(p)\,y_2(q)\Big]= \\ \nn
&& -\frac{i}{16\pi\,g^2}\frac{\big(1-p_-^2\big)^2\big(1-q_-^2\big)^2}{p_- q_-\big(p_-+q_-\big)^2}\log \frac{q_-}{p_-}
+\frac{i}{32\pi\,g^2}\frac{(1-p_-^2)(1-q_-^2)(p_--q_-)\big(1+p_-q_-\big)^2}{(p_-q_-)^2\big(p_-+q_-\big)}
\eea
while the second equals
\bea
&&  i \, \text{Im }\mathcal{A}^{(1)}\Big[y_1(p)\, \bar y_2(q)\rightarrow  y_1(p)\,\bar y_2(q)\Big]= \\ \nn
&& -\frac{i}{16\pi\,g^2}\frac{\big(1-p_-^2\big)^2\big(1-q_-^2\big)^2}{p_- q_-\big(p_--q_-\big)^2}\log \frac{q_-}{p_-}
-\frac{i}{32\pi\,g^2}\frac{(1-p_-^2)(1-q_-^2)(p_-+q_-)\big(1+p_-q_-\big)^2}{(p_-q_-)^2\big(p_--q_-\big)},
\eea
where we for simplicity chose to write the expressions in terms of right moving momenta alone. It is interesting to note that all finite terms of the amplitude solely originate from the quartic interactions. The only role of the six-vertex tadpoles is to make the final amplitude UV finite.

The one-loop BOSST phase can be written as an infinite sum \cite{Borsato:2013hoa,Abbott:2013ixa},
\bea \label{eq:BOSST}
&& \theta_{LL}=-\frac{i}{\pi}\sum_{r=1}^\infty \sum_{s=r+1}^\infty \Big[\frac{1-(-1)^{r+s}}{2}\Big(\frac{s-r}{r+s-2}-\frac{1}{2}\big(\delta_{r,1}-\delta_{s,1}\big)\Big)\big(x^r y^s-x^sy^r\big)\Big]
\\ \nn
&& \tilde\theta_{LR}=-\frac{i}{\pi}\sum_{r=1}^\infty \sum_{s=r+1}^\infty \Big[\frac{1-(-1)^{r+s}}{2}\Big(\frac{r+s-2}{s-r}-\frac{1}{2}\big(\delta_{r,1}-\delta_{s,1}\big)\Big)\big(x^r y^s-x^sy^r\big)\Big]
\eea
where
\bea \nn
x^n=\frac{p_1}{h}\big(\frac{\omega_p-1}{p_1}\big)^{n-1}+\mathcal{O}(h^{-3}),\qquad
y^n=\frac{q_1}{h}\big(\frac{\omega_q-1}{q_1}\big)^{n-1}+\mathcal{O}(h^{-3})
\eea
and $\omega$ denotes the relativistic dispersion relation with momenta $q_1$ and $p_1$. The Kronecker-deltas in (\ref{eq:BOSST}) are the novel terms in the modification \cite{Borsato:2013hoa,Abbott:2013ixa} and are, as mentioned above, only visible in the strict near-BMN.

Evaluating the sum in (\ref{eq:BOSST}) and substituting the explicit expressions for $x$ and $y$ it is easy to verify that we have a perfect match upon the identification $h=2g$. Of course, if we were to simply ignore the novel Kronecker-terms, then the phase would not agree with the worldsheet result.
\section{Conclusions and summary}
We have investigated two and four-point functions at the one-loop level. The two-point functions allowed us to probe the exact dispersion relation for the light modes, which is fixed by integrability up to an interpolating scalar function. The scalar function made its appearance already for the more symmetrical $AdS_4 / CFT_3$ where its apparent regularization dependence caused some controversy in the literature \cite{Alday:2009zz,Krishnan:2008zs,McLoughlin:2008he,McLoughlin:2008ms,Gromov:2008fy,Astolfi:2011ju}. Depending on whether the heavy mode of the worldsheet theory is treated as a composite state or not, different one-loop predictions is obtained. The same situation appears also for the less symmetric $AdS_3 / CFT_2$ \cite{Forini:2012bb,Sundin:2012gc}. Standard QFT lore dictates that a regulator should be chosen so that none of the underlying symmetries of the problem are broken. By using methods from implicit regularization we demonstrated that only the WS-regulator (which treats all worldsheet fields equally) was consistent with worldsheet supersymmetry. This then implies that the interpolating scalar function indeed exhibits a non-zero one-loop term.

We then turned to an investigation of the one-loop phase. The one-loop part of the phase was first proposed in \cite{Beccaria:2012kb} and later modified with additional terms in \cite{Borsato:2013hoa,Abbott:2013ixa}. By utilizing the full BMN string we computed $2\rightarrow 2 $ scatterings on the worldsheet and confirmed that these agreed with the modified one-loop phase.

There are several possible continuation of the work presented in this letter. For example, it would be interesting to perform the two-point function analysis in the setting of $AdS_4\times \mathbbm{CP}_3$ using the methods described in this paper. It is plausible that the same result would be found but it would nevertheless be interesting to verify this explicitly. Furthermore, it would also be interesting to probe the $S^3\times S^1$ geometry by extending the scattering analysis beyond $\alpha=0,1$. However, this would involve a much larger class of Feynman topologies and the explicit computation would probably be fairly involved, at least if the full BMN string is considered.
\section*{Acknowledgements}
It is a pleasure to thank Michael Abbott, Silvia Penati, Jonathan Shock  and Linus Wulff for valuable discussions. I would also like to extend my gratitude to Michael Abbott and Linus Wulff for comments and suggestions on the manuscript. This work was supported by a joint INFN and Milano-Bicocca postdoctoral grant.
\bibliographystyle{JHEP}
\bibliography{2and4point_v1}
\end{document}